# Super-resolution deep imaging with hollow Bessel beam STED microscopy

Wentao Yu[1], Ziheng Ji[1], Dashan Dong[1], Xusan Yang[2], Yunfeng Xiao[1, 3], Qihuang Gong[1, 3], Peng Xi[2*], and Kebin Shi[1, 3*]

[1]State Key Laboratory for Mesoscopic Physics, Collaborative Innovation Center of Quantum Matter, School of Physics, Peking University, Beijing 100871, China
[2]Department of Biomedical Engineering, College of Engineering, Peking University, Beijing 100871, China
[3]Collaborative Innovation Center of Extreme Optics, Shanxi University, Taiyuan, Shanxi 030006, China
*E-mail: kebinshi@pku.edu.cn  xipeng@pku.edu.cn

*Abstract*—Stimulated emission depletion (STED) microscopy has become a powerful imaging and localized excitation method breaking the diffraction barrier for improved lateral spatial resolution in cellular imaging, lithography, etc. Due to specimen-induced aberrations and scattering distortion, it is a great challenge for STED to maintain consistent lateral resolution deeply inside the specimens. Here we report on a deep imaging STED microscopy by using Gaussian beam for excitation and hollow Bessel beam for depletion (GB-STED). The proposed scheme shows the improved imaging depth up to ~155μm in solid agarose sample, 115μm in PDMS and ~100μm in phantom of gray matter in brain tissue with consistent super resolution, while the standard STED microscopy shows a significantly reduced lateral resolution at the same imaging depth. The result indicates the excellent imaging penetration capability of GB-STED pave its way for deep tissue super-resolution imaging and 3-D precise laser fabrication.

*Keywords:* stimulated emission depletion microscope, Bessel beam, super resolution, deep imaging

## I. INTRODUCTION

STIMULATED emission depletion (STED) optical microscopy breaks the barrier of Abbe diffraction limit, enabling the visualization of the subcellular structure and dynamics by suppressing the spontaneous emission selectively.[1-4] Typically, a doughnut-shaped depletion beam surrounding the excitation focus is formed via helical phase modulation. The designated beam profile leads to a highly confined fluorescent emitting area beyond the diffraction limit by deactivating the fluorophores in the annulus surrounding by stimulated emission.[5-7] Therefore, the fluorescent signal can be only generated from the central emitting area, leading to a narrow point spread function (PSF). Various modalities possessing spatial resolution beyond the diffraction limit[8-11] based on STED mechanism have been developed recently and found important applications in live-cell nanoscope,[12-15] laser fabrication,[16, 17] and nano-structure characterizations.[6, 18, 19]

The achievable resolution in STED systems explicitly relies on the fluorophore depletion efficiency, i.e. the efficient use of photons delivered by the designated depletion beam. As a result, it is challenging to maintain consistent resolution as the foci moves from the surface to deep inside of the specimens due to the spherical aberration, scattering distortion and loss.[12, 15, 20] Experimental and theoretical efforts have been reported to improve the deep imaging resolution in STED based modalities such as utilizing adaptive optics,[21] compensating spherical aberration with correction collar[14] and polarization manipulation.[22] A sustainable super resolution is demonstrated experimentally at the maximum depth of 80~100μm previously.[14] Alternatively, it is noticeable in light sheet microscopy that a system less susceptible to scattering distortions could be achieved by using Bessel beam.[23] Similarly, line scanned Bessel beam STED (BB-STED) is theoretically proposed by using Bessel beam in both excitation and depletion beams in light sheet microscopy, for improved sectioning thickness and reduced photo-damage. [24]

In this Letter, we report on the demonstration of a maintainable resolution at deep imaging depth with Gaussian-Bessel STED (GB-STED) microscopy. We modulate the depletion beam into a hollow Bessel beam and use a conventional Gaussian beam as the excitation. Taking advantage of the distortion free nature of the Bessel beam, the doughnut structure is maintained at an extended distance in focal region. Further, it is advantageous to generate a Gaussian PSF excitation with confined axial distribution, as modulating the excitation beam to a Bessel form generates unnecessary excitation outside the axial detection region, which induces severe out-of-focus noise and photo-bleaching. Through the GB-STED scheme, lateral resolution of deep imaging was observed barely un-changed at the maximum depth of 155μm in solid agarose sample and 97μm in agarose phantom of brain tissue in our experiment, while the standard STED microscopy shows a significantly degraded lateral resolution at the same sample depth in our experiments.

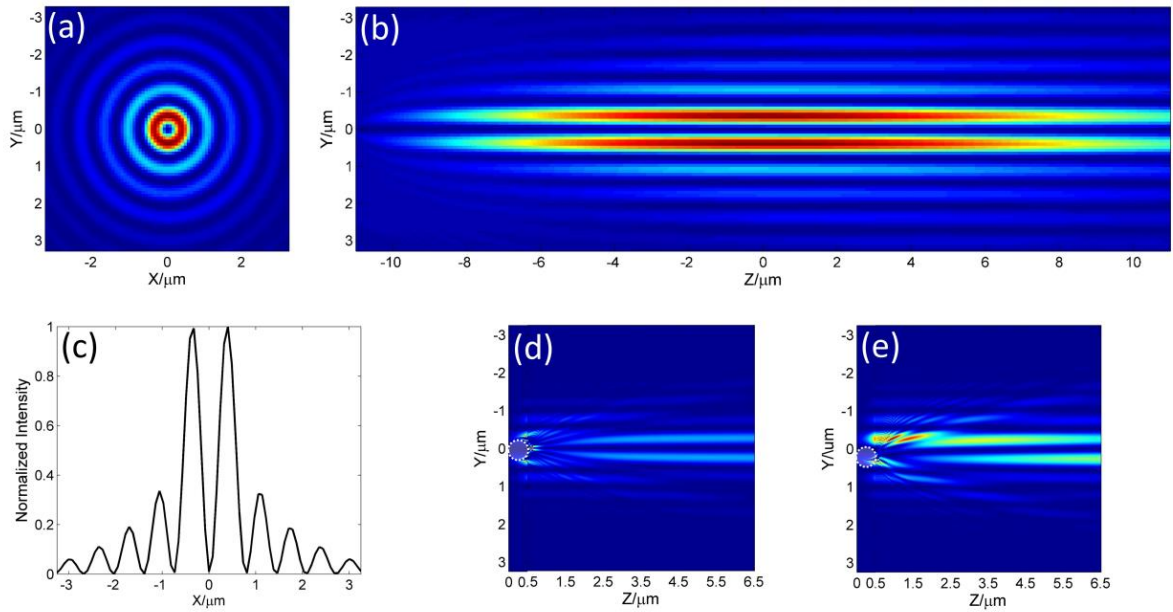

Figure 1. Intensity distributions of the hollow Bessel beam: (a) in the x-y plane at z=0; (b) in the x-z plane; (c) cross section profiles along x-axis; (d)-(e) scattered by a 500nm-diameter bead (d) on the optical axial at z=0, (e) 250 nm off the optical axial at z=0.

## II. SCHEMATIC DIAGRAM OF GB-STED

To investigate the non-diffraction and self-healing properties of the hollow Bessel beam, we simulate the beam generation and scattering behaviors based on scalar diffraction theory and fast Fourier transforms beam-propagation method (FFT-BPM).[25] A Gaussian incident beam ($\lambda = 750$ nm) can be described as $E(x, y, z) = E_0/\omega_0 * \exp[(x^2 + y^2)/\omega_0^2]$ at the waist plane, where $E_0$ is the field amplitude and $\omega_0 = 8\ mm$ is defined as the radius of the beam waist. The beam transmits through a vortex 0 to 2π phase plate described as $\tilde{T} = \exp(i\varphi)$ and a followed axicon lens (refractive index = 1.4542) with the apex angle of 176 degree to generate a hollow first-order Bessel beam. Then a telescope system composed of a lens (f = 200mm) and an objective lens (f = 2mm) relays the hollow Bessel beam to the foci region of the objective lens.

**Figure 1**(a) and (b) show the simulated intensity distributions of the hollow Bessel beam in the x-y plane at $z = 0$ (the focal plane of the objective lens) and in the x-z plane respectively. An extended depth of field can be observed in the Bessel zone shown in Figure 1(b) as expected. It can also be observed that the transverse profile of the hollow Bessel beam exhibits the zero-intensity at the center surrounded by successive rings as shown in Figure 1(c), a feature which allows its use as the depletion beam.

We further investigated the influence of scatter center towards the incoming hollow Bessel beam. In our simulation, a virtual micro-bead with 500 nm diameter and refractive index of 1.59 was immersed in the medium with index of 1.33 and placed at different transverse locations within the beam profile. The simulated results are shown in Figure 1(d) and (e), where Figure 1(d) shows the field intensity distributions in x-z plane when the bead is located on the optical axis at $z = 0$ and Figure 1(e) shows the distribution when the bead is located 250 nm off the optical axis at $z = 0$. The reconstruction of hollow Bessel beam immediately after the bead exhibits noticeable rotational distortion when the bead is located off-axis as shown in Figure 1(e), since orbital angular momentum induced by the helical phase modulation exhibits rotational amplitude distribution when encountering the asymmetric scattering.[26, 27] It is therefore different from the observation in conventional Bessel beam scattering behavior discussed previously.[28] As shown in Figure 1(e), the doughnut-shaped beam can be recovered after few micrometer propagation distance, where the STED imaging can be still performed.

The schematic diagram of our GB-STED imaging system is shown in **Figure 2**(a). An output beam of 750 nm wavelength from a CW Ti:Sapphire laser (M Squared, SolsTiS SRX-F-10) and a diode CW laser of 635 nm wavelength (Boson Technology, DL-030-1) were used as depletion beam and excitation beam, respectively. The depletion beam was converted into hollow Bessel beam after passing through a vortex phase plate (RPC Photonics, VPP-1a) and an axicon lens (Altechna, 1-APX-2G254-P) with the apex angle of 176 degree. The generated hollow Bessel beam was relayed onto the sample by a telescope system consisted of lens L1 and the final objective lens (Olympus, PL APO 60X /NA=1.2 water immersion) as shown in Figure 2(a). The fluorescence excitation beam was first spatially filtered, and then focused onto the sample. Two collimated beams were combined with DM1 and DM2 (Chroma, ZT640rdc and ZT740sprdc). A symmetrical depletion beam distribution with a nearly zero-intensity center (3.7% of the peak intensity of Bessel doughnut PSF, shown in **Figure S1**) and an ideal isotropic excitation were obtained by imposing circular polarizations to both depletion beam and excitation beam in the back-aperture of the objective lens using a half-wave plate and a quarter-wave plate (Thorlabs, AHWP05M-600 and AQWP05M-600). The fluorescence signal was collected by the same objective lens, passed through a band-pass filter (Chroma, HQ675/50M) before being focused into a 50μm-diameter multimode fiber (Thorlabs, FG050UGA) as the confocal detection pinhole (1.04 Airy Unit) connected to a photon-counting avalanched photon diode (Perkin Elmer, SPCM-AQRH-13-FC). The inset in Figure 2(a) shows the magnified view of the light field after the objective lens. In addition, the shown GB-STED imaging system can be readily converted to conventional STED microscope by translating out both the axicon lens and lens L1.

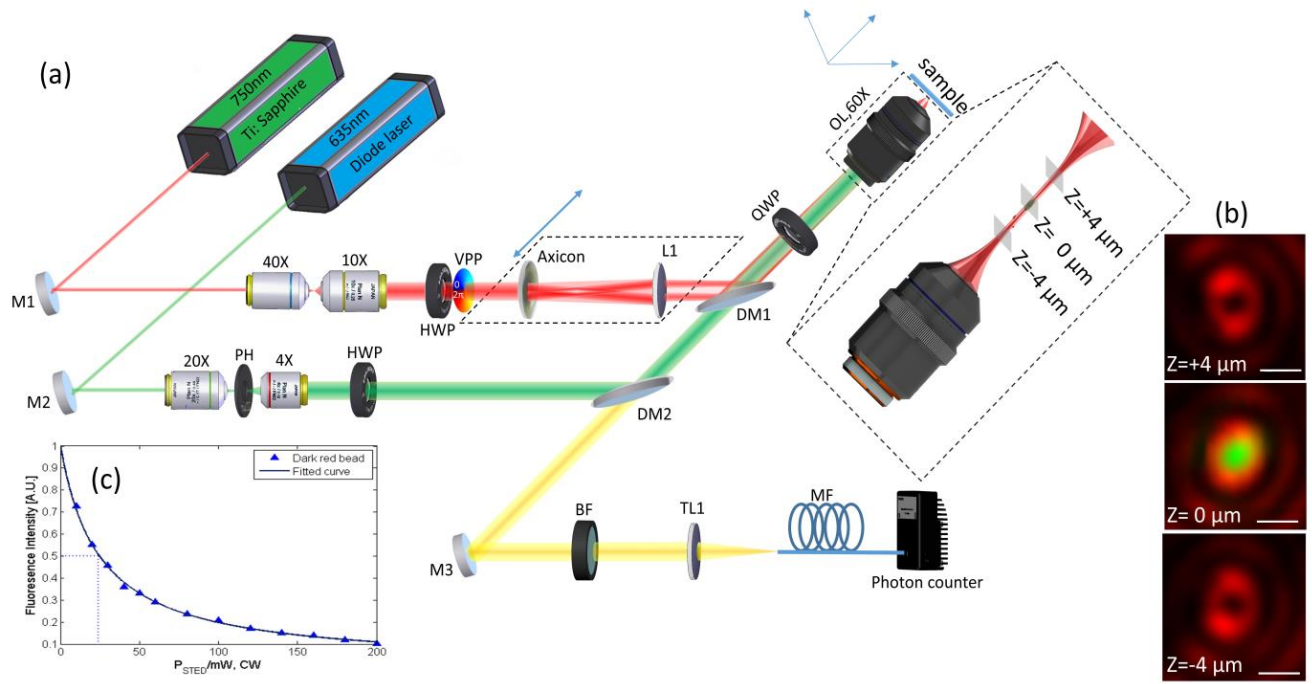

Figure 2. (a) Schematic diagram of the GB-STED setup. L1: singlet lens, focal length 200 mm; TL1: tube lens, focal length 200 mm; OL: objective lens; M1, M2, M3: silver mirror; HWP: half-waveplate; QWP: quarter-waveplate; PH: pinhole, diameter 10 μm; VPP: vortex phaseplate; DM1,DM2: dichroic mirror; BF: band-pass filter; MF: multimode fiber. The inset shows magnified view of the boxed area. (b) Measured beam profile of excitation beam and depletion beam at z=0,±4μm, where z=0 means the focal plane of the objective lens. Scale bar: 500 nm. (c) Fluoresence intensity versus the depletion power was measured with dark red beads.

We obtained the beam profile at the foci region for both excitation beam and depletion beam, as shown in Figure 2(b), by mechanically scanning an 80nm-diameter golden particle (BBI Solution) in the lateral plane at $z = 0, \pm 4\mu m$ with a piezo stage (Thorlabs, MAX311D/M). The golden bead can back-scatter both Gauss excitation beam and doughnut-shaped depletion beam, hence both profiles of excitation and depletion can be obtained. On the contrary, the fluorescent bead can only be excited by the excitation beam, therefore it cannot be used for probing the PSF profile. Scattered signal was collected via a non-confocal configuration without the band-pass filter. As shown in Figure 2(b), hollow Bessel depletion beam was effectively generated for an extended distance. More detailed illustration for beam profile can be found in Support Information Section 1. The STED imaging acquisition was carried at z = 0, where depletion beam and excitation beam were overlapped. The lateral resolution of our setup is determined by the effective fluorescence emitting area suppressed by the doughnut-shaped depletion beam, and the axial resolution by the confocal configuration.

III. RESULTS

To test the performance of GB-STED at different imaging depth, dark red fluorescent beads of 40nm-diameter (Invitrogen) were used in our experiments. To estimate the resolution of the STED system, the saturation power $P_s$ (the depletion power at which the fluorescence intensity drops to half) of the fluorescent specimen was measured by overlapping the excitation beam and a regularly focused Gaussian depletion beam without phase modulation at foci. [29, 30] Then a layer of fluorescence beads air dried and mounted in water on a cover glass was used as the sample. The fluorescence signal integrated over the field of view was recorded along with different depletion powers[6] as shown in Figure 2(c). The fitting of $\eta = 1/(1 + P_{STED}/P_{sat})$ agrees well with the experimental data; and we obtain the saturation power- $P_s = 24.63 mW$. For a STED beam of 200mW, the theoretical resolution of conventional STED calculated by resolution = $\lambda exc/[2NA(1 + P_{STED}/P_{sat})]$,[31] where $\lambda exc$ = 635nm and effective NA = 1.01 (see discussion in Support Information Section 1 for effective NA used), is about 103nm.

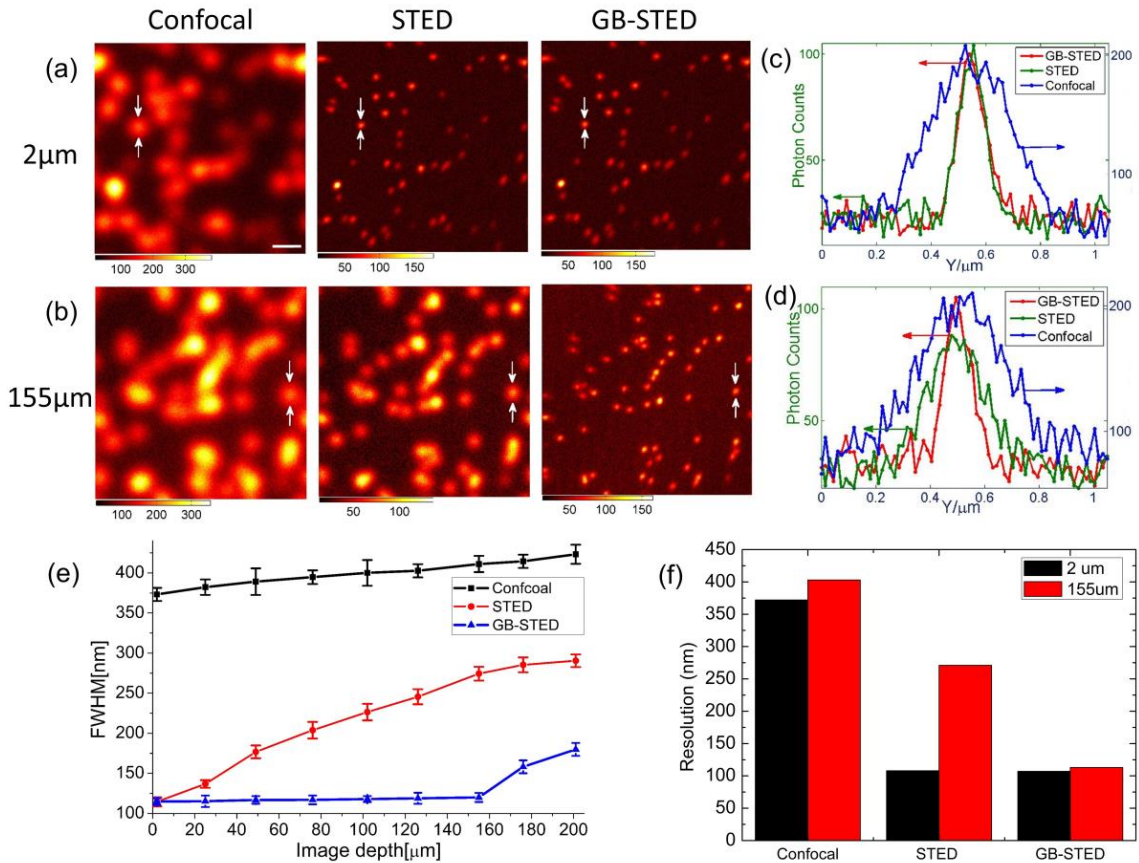

Figure 3. Imaging of dispersed 40 nm fluorescence beads in solid agarose sample via confocal, conventional STED and GB-STED. (a) Imaging acquired at the sample surface (2 μm depth) via confocal, conventional STED and GB-STED; (b) imaging acquired at the sample depth of 155 μm via confocal, conventional STED and GB-STED. (c) Cross section of selected bead in (a); (d) cross section of selected beam in (b). (e) Averaged FWHM of the 40nm-diameter beads at different depth of confocal, conventional STED, and GB-STED imaging. (f) Resolution of confocal, conventional STED and GB-STED at 2 μm and 155 μm. Scale bar: 500 nm. Dwell time: 0.4 ms

## A. Deep imaging with agarose-embedded fluorescent beads

For deep imaging experiments, the 40nm fluorescent beads were diluted in 70 ℃ water mixed with low temperature melting point gel (Lifetechnologies, UltraPure™ Low Melting Point Agarose) and then solidified at room temperature. With a STED beam of 200mW and excitation beam of 40μW in the back-aperture of the objective lens, confocal, STED and GB-STED imaging were carried out at the surface (2μm in depth) of the sample and 155μm in depth of the sample respectively. As shown in **Figure 3**(a), at the surface of the sample, the full-width at half-maximum (FWHM) of the fluorescent nanoparticle with conventional STED (115nm) and GB-STED (114nm) are almost modalities. After deconvolution of the size of the fluorescent beads, the resolutions of both modalities (108 nm) are in good agreement with the theoretical prediction of 103 nm. This is almost 3.5-fold improvement on resolution comparing to confocal imaging as shown in Figure 3(c) and (f). When imaging deep inside the specimen, the aberration and scattering may greatly alter the zero-intensity of the center of conventional STED PSF, resulting in much worse resolution. We took images up to ~200μm deep in the sample with depth increment of ~25μm. The optical resolution of conventional STED decreases monotonically with the depth as shown in red in Figure 3(e). When imaging at 155μm, the measured resolution for conventional STED dropped to 271nm, whereas GB-STED can maintain 113nm resolution and unchanged signal counting reads, as shown in Figure 3(b), (d) and (f).

Figure 3(e) shows the retrieved FWHM averaged over 10 beads as a function of acquisition depth for different imaging modes. At depths beyond 155μm the resolution of GB-STED deteriorated quickly as well due to increased scattering distortion and depletion power loss. Despite this, GB-STED exhibits consistent resolution up to about 155μm which is insusceptible to the distortions produced by scattering and index mismatching, while conventional STED has seen a 2.5-fold decrease in its lateral resolution at the same depth. The measured profile of depletion beam of traditional STED and GB-STED, shown in **Figure S3,** also support the fact that traditional STED pattern deteriorates with the imaging depth increasing, while GB-STED pattern keeps nearly invariant.

## B. Deep imaging with refractive index mismatch

Unlike the agarose samples, most bio-samples are with relatively large refractive index mismatch. To demonstrate the performance of GB-STED deep imaging with large aberrations induced by refractive index mismatch, we also imaged 40nm-diameter beads embedded in polydimethylsiloxane (PDMS, refractive index: 1.41) at different depth and compared the result of conventional STED with objective collar correction to compensate for the aberration at different depth. GB-STED exhibits consistent super resolution up to 115μm in depth, while standard STED has seen a 2.3-fold decreasing in its lateral resolution at the same depth. The detailed results and discussions are included in Support Information Section 2.

*C. Deep imaging with gray matter phantom*

To demonstrate the potential of GB-STED microscopy in deep tissue imaging with strong scattering and absorption, we have prepared a phantom (details in Support Information Section 3) with extinction coefficients similar to that of the gray matter of the brain tissue [32], with 40 nm fluorescent beads embedded. The absorption and reduced scattering coefficients of agarose phantom are $0.424 cm^{-1}$ and $10.236 cm^{-1}$ (shown in **Figure S6**), which are similar to gray matter [32]. With the same excited and depletion power, the optical resolution of conventional STED dropped by half (from 108 nm to 217nm) at the sample depth of 97μm, whereas the resolution of GB-STED remains almost unchanged (from 109nm to 112nm). As a matter of fact, for our phantom sample, shown in **Figure 4**, GB-STED microscopy exhibits consistent resolution up to about 100μm depth as shown in Figure 4(e).

IV. CONCULSIONS

In summary, we have reported on a Gaussian-Bessel beam STED imaging modality experimentally. The proposed scheme shows the capability of maintaining super resolution deep inside of the sample up to ~155μm in solid agarose sample and ~100μm in phantom similar to gray matter in brain tissue, owing to advantages of non-diffractive and self-reconstructing natures of the hollow Bessel depletion beam. Benefiting from the Gaussian PSF excitation, the photobleaching effect is minimized. We have demonstrated the capability of GB-STED in deep imaging in a highly turbid medium with large refractive index mismatch conditions with CW lasers. Other temporal modulation techniques, such as pulsed STED and time-gated STED can be readily implemented to further improve the imaging quality.[33] Comparing with other active phase modulation methods such as collar correction and adaptive optics, GB-STED can achieve deep imaging directly, without any intervene, which makes GB-STED a fast and robust 3D super-resolution technique. It can also provide a promising platform for super-resolution laser nano-fabrication modalities

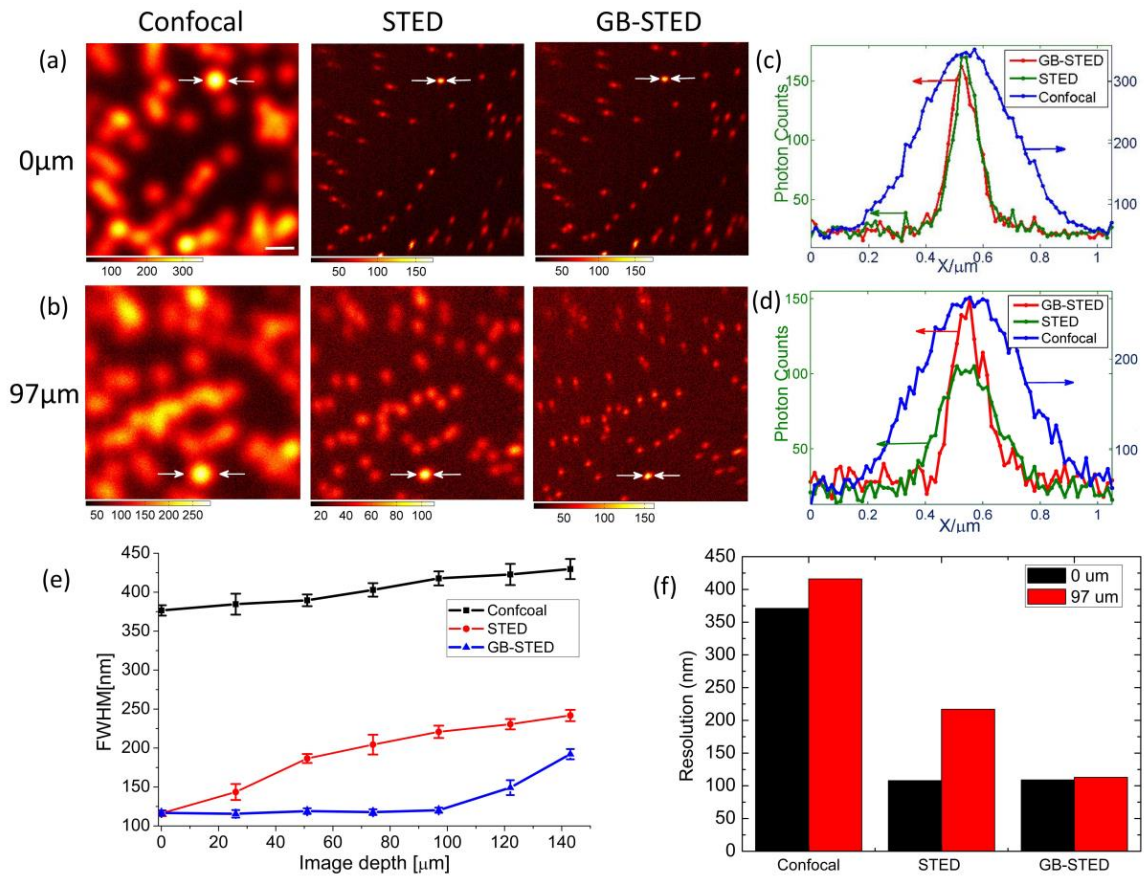

Figure 4. Imaging of dispersed 40 nm fluorescence beads in agarose phantom via confocal, conventional STED and GB-STED (a) Imaging acquired at the sample surface (0 μm depth) via confocal, conventional STED and GB-STED; (b) imaging acquired at the sample depth of 97 μm via confocal, conventional STED and GB-STED. (c) Cross section of selected bead in (a); (d) cross section of selected beam in (b). (e) Averaged FWHM of the 40nm-diameter beads at different depth of confocal, conventional STED, and GB-STED imaging. (f) Resolution of confocal, conventional STED and GB-STED at 0 μm and 97 μm. Scale bar: 500 nm. Dwell time: 0.4 ms

such as STED based optical data storage [17] where super-resolution laser writing and reading in deep volume are desired.

**Acknowledgements** We acknowledge the funding support from the National Science Foundation of China (NSFC#11174019, 61322509, 61475010, and 11121091) and the Ministry of Science and Technology of China (National Basic Research Program of China under Grant No. 2013CB921904). Authors thank Xuanze Chen and Yujie Sun for fruitful discussions and helps in sample preparation.

Supporting Information

**Title: Super resolution deep imaging by using hollow Bessel beam STED microscope**

*Wentao Yu[1], Ziheng Ji[1], Dashan Dong[1], Xusan Yang[2], Yunfeng Xiao[1, 3], Qihuang Gong[1, 3], Peng Xi[2*] and Kebin Shi[1, 3*]*

[1]State Key Laboratory for Mesoscopic Physics, School of Physics, Peking University, Beijing 100871, China
[2]Department of Biomedical Engineering, College of Engineering, Peking University, Beijing 100871, China
[3]Collaborative Innovation Center of Quantum Matter, Beijing 100871 China

1. **Characterization of system point spread function and numerical aperture**

The system point spread functions (PSFs) for different imaging modalities were measured by mechanically translating a single golden nano-particle of 80nm diameter in the focal plane. The nano-particle was immobilized on a #1 microscope cover glass. The golden bead can back-scatter both Gauss excitation beam and doughnut-shaped depletion beam, hence both profiles of excitation and depletion can be obtained. On the contrary, the fluorescent bead can only be excited by the excitation beam, therefore it cannot be used for probing the PSF profile. The scattered signal was collected by a photon counter as shown in Figure 2 (a) with multimode fiber removed, i.e. in a non-confocal collection mode. Figure S1 shows the cross section profiles for hollow Bessel depletion beam and Gaussian excitation beam respectively at the imaging focal plan in GB-STED scheme. The excitation beam is well confined at the center of doughnut pattern without any overlapping with side lobes. The minimum intensity at the center of hollow Bessel depletion beam is measured 3.7% of the maximum intensity, indicating a functional depletion beam for STED imaging.

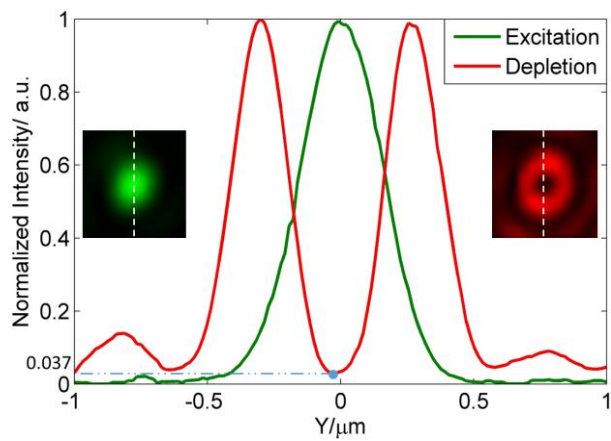

**Figure S1.** Cross section of Gaussian excitation beam (green) and hollow Bessel depletion beam (red)

By removing the axicon and L1 shown in Figure 1(a) from the beam path, we could convert the system to a conventional STED microscope. Figure S2 shows the PSFs measured at the objective lens focal plane in a conventional STED. A normal doughnut shaped depletion pattern without side lobes can be observed. The full width half maximum (FWHM) of Gaussian excitation beam is measured at 383nm, indicating an effective numerical aperture (NA) of 1.01 based on the equation FWHM=0.61*wavelength/NA. For depletion beam, the peak-to-peak distance of the doughnut shape is measured at 738nm, indicating an effective NA of 1.016 by using $D_{peak-to-peak}$= wavelength/NA. [1] The measured effective NA was used for calculating the achievable lateral resolution for the STED and GB-STED imaging systems.

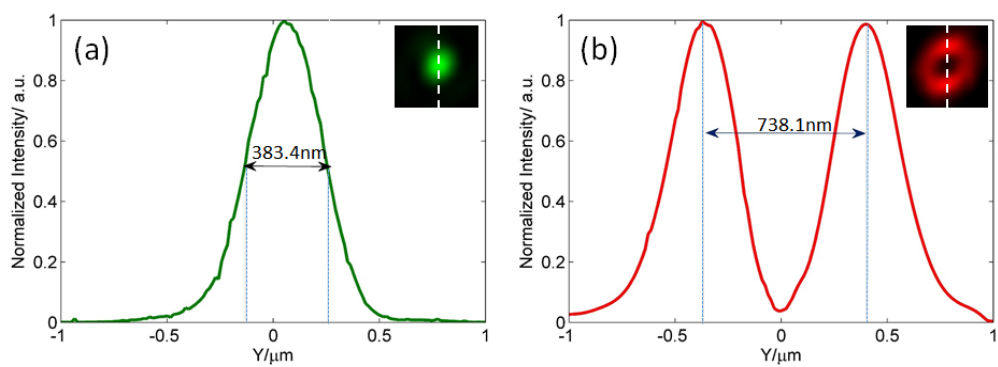

**Figure S2.** Cross section of Gaussian excitation beam (green) and conventional doughnut shaped depletion beam (red)

The degradation of resolution in STED microscope for deep sample imaging is mainly due to the distortion of the doughnut shaped depletion beam. [2-4] In order to better understand the mechanism, it is critical to obtain the depletion PSFs *in situ* along with the deep-imaging acquisition process. We prepared the mixture of diluted golden nanoparticles and 40nm-diameter fluorescence beads in solid agarose sample. Therefore the beam profile measurement and the imaging acquisition can be performed by using the golden particle and fluorescence bead in the mixture respectively. In the experiments, PSF measurements on depletion beam were carried at the same sample depth where the imaging of fluorescent bead was performed. The results are shown in Figure S3. As shown in the insets in Figure S3 (a)-(c), the depletion beam gets distorted with the imaging depth for conventional STED setup while GB-STED pattern keeps nearly invariant as shown in Figure S3 (d)-(f). Significant amount of background floor can be observed in the center region of the doughnut shaped profile for conventional STED. Since the resolution of STED microscope is very sensitive to the depletion beam profile, a degradation of resolution is resulted at deep imaging depth in conventional STED microscope. Considering the bead size, the optical resolutions at the depth of 2μm are 108 nm for STED and 109 nm for GB-STED, which are almost same. But at the depth of 76μm, the resolution of STED worsen to 197 nm, while the resolution of

GB-STED keep nearly constant ~111nm. Even at depth of 155μm, the resolution is improved from 271 nm (STED) to 112nm (GB-STED).

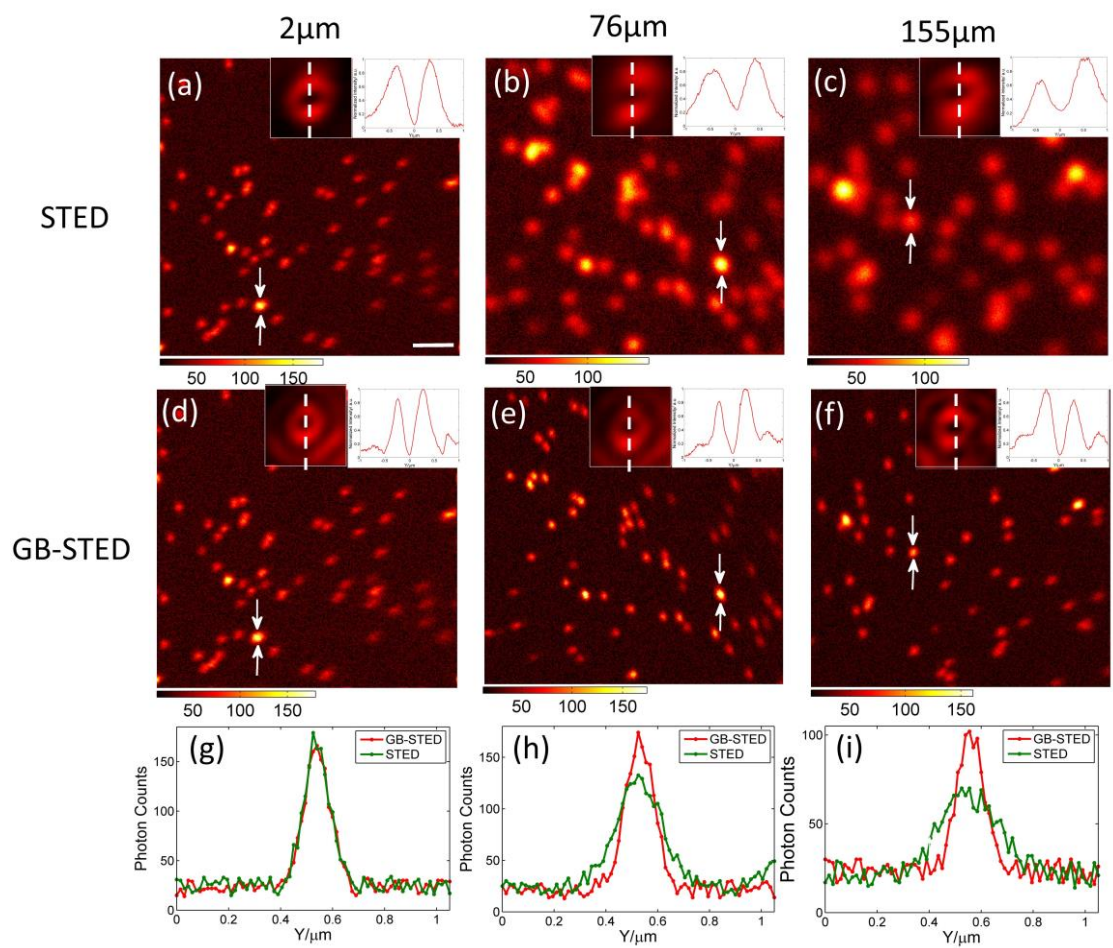

**Figure S3.** Lateral PSF measurements of depletion beam and imaging of dispersed 40 nm fluorescence beads in solid agarose samples via standard STED and GB-STED (Excited beam: 40μW, Depletion beam: 200mW). (a)-(c) Imaging via conventional STED at different depth of 2μm, 76μm and 155μm. (d)-(f) Imaging via GB-STED at different depth of 2μm, 76μm and 155μm. The insets in (a)-(f): Measured beam profile of depletion beam and cross section along y-axis at each depth. (g) Cross section of selected bead in (a) and (d), FWHM: 115 nm(conventional STED), 116 nm(GB-STED); (h) cross section of selected bead in (b) and (e), FWHM: 201 nm(conventional STED), 118 nm(GB-STED); (i) cross section of selected bead in (c) and (f), FWHM: 274 nm(conventional STED), 119 nm(GB-STED). Scale bar: 500nm. Dwell time: 0.4ms

## 2. Imaging in sample with large refractive index mismatch

The aberration resulting from refractive index mismatch plays critical role in STED super-resolution microscopy, because it strongly affects the distribution of the STED doughnut PSF. Previously for thick specimen STED imaging, there are two methods to solve this issue. The first is by using a glycerol-immersion lens to best match the refractive index between the immersion fluid and the tissue, and to adjust the correction collar of the objective [2, 3, 5]. Depending on the experimental refractive index condition, the resolution can be maintained for imaging up to 100μm [5], while sometimes the resolution at 90μm depth is already 2.5x worse than that at surface [3]. The second method is to use adaptive optics to compensate for the aberrations induced [6], with imaging 25μm depth retina tissue were demonstrated.

Here, we show that with GB-STED, the super-resolution can be maintained with over ~100μm depth, at a large refractive index mismatch scenario. To make a large mismatch of refractive index between samples and water immersed lens used, we prepared polydimethylsiloxane (PDMS, refractive index: 1.41) based sample containing 40nm-diameter fluorescence beads to measure the resolution of confocal, conventional STED and GB-STED at different depth. The imaging results are shown in Figure S4. Considering the bead size, the optical resolutions of confocal, conventional STED and GB-STED are 371 nm, 107 nm and 109 nm at the depth of 1μm. At the depth of 115μm, the resolutions are measured at 431 nm, 242 nm and 118nm respectively for confocal, STED and GB-STED. As shown in Figure S4 (e) and (f), GB-STED exhibits consistent resolution up to 115μm, while standard STED has seen a 2.3-fold decreasing in its lateral resolution at the same depth.

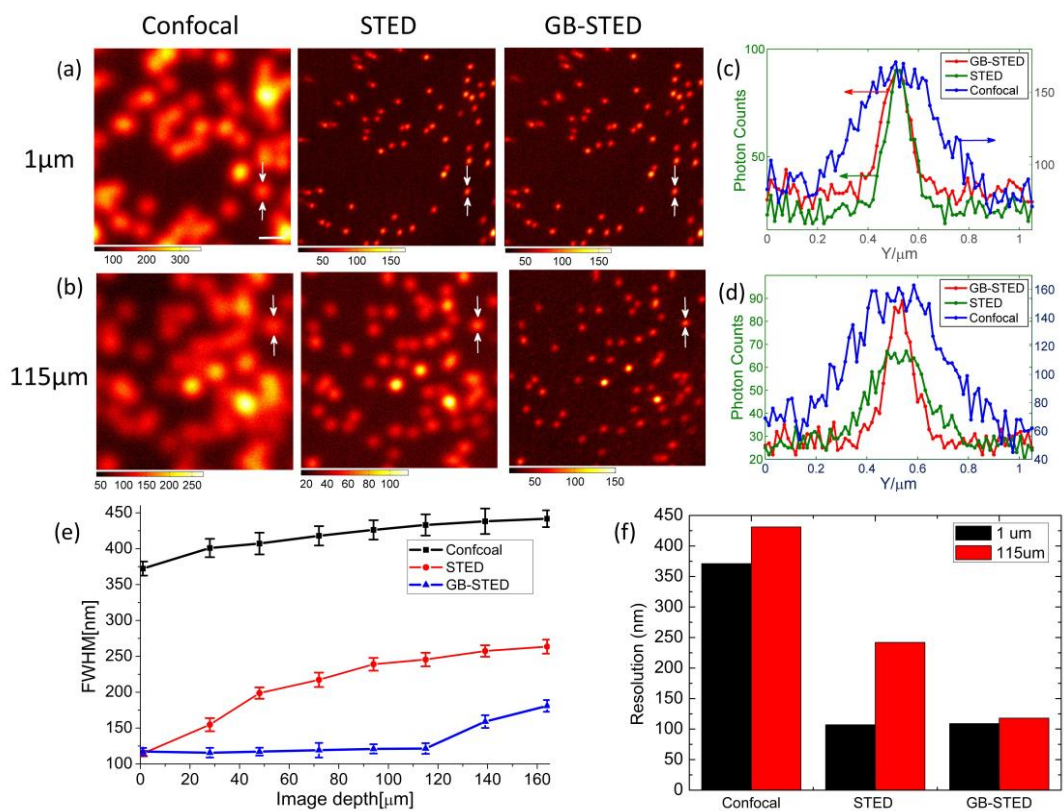

**Figure S4.** Imaging of dispersed 40 nm fluorescence beads in PDMS (refractive index of 1.41) via confocal, standard STED and GB-STED (Excited beam: 40μW, Depletion beam: 200mW). (a) Imaging acquired at the sample surface (1μm depth) for confocal, standard STED and GB-STED; (b) imaging acquired at the sample depth of 115μm for confocal, standard STED and GB-STED. (c) Cross section of selected bead in (a) for different modalities respectively, FWHM: 373 nm (Confocal), 114 nm (conventional

STED), 116 nm(GB-STED); (d) cross section of selected bead in (b) for different modalities respectively, FWHM: 433 nm (Confocal), 246 nm (conventional STED), 125 nm(GB-STED). (e) Averaged FWHM of the 40nm-diameter beads at different depth of confocal, standard STED, and GB-STED imaging. (f) Resolution of confocal, conventional STED and GB-STED at sample depth of 1μm and 115μm. Scale bar: 500 nm. Dwell time: 0.4ms.

We further investigated the comparison between GB-STED method and conventional STED with correction collar. We prepared the sample by embedding mixture of 80nm golden particles and 40 nm fluorescent beads into PDMS (refractive index of 1.41). With a water-immersion objective (refractive index of 1.33), the STED imaging with/without adjusting the correction collar is performed, and the result is compared with GB-STED counterparts. The PSF of depletion beam under different schemes were also obtained *in situ* through non-confocal scattering imaging of golden nanoparticle. As can be seen from Figure S5, without collar correction, the resolution at 46μm depth reaches 186 nm, and it reaches 214 nm at 94μm depth. For conventional STED, after collar correction, the resolutions are improved from 186 nm to 110 nm (at 46μm depth) and 214 nm to 147 nm (at 94μm depth), respectively. While for GB-STED, the resolution are kept nearly constant (109 nm at 46μm and 114nm at 94μm). This clearly shows the advantage of using GB-STED for deep tissue imaging. Moreover, unlike the collar correction which is time consuming, iterative, and empirical [5], GB-STED does not need any adjustment during the imaging process.

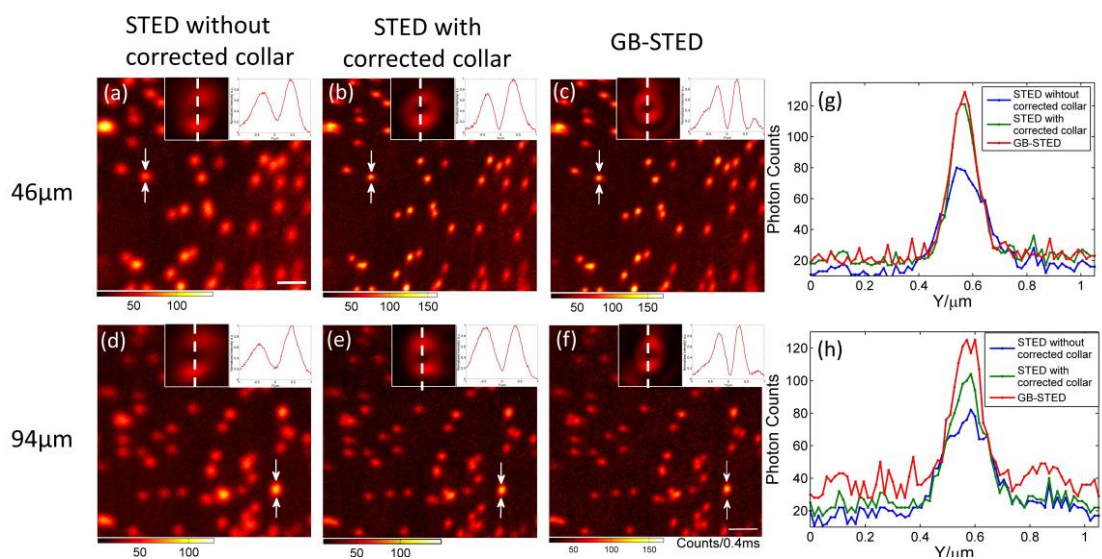

**Figure S5.** Lateral PSF measurements of depletion beam and imaging of dispersed 40 nm fluorescence beads in PDMS via STED without corrected collar, STED with corrected collar and GB-STED (Excited beam: 40μW, Depletion beam: 200mW). (a)-(c) Imaging at depth of 46μm: (a) STED without corrected collar, (b) STED with corrected collar, (c) GB-STED. (d)-(f) Imaging at depth of 94μm: (d) STED without corrected collar, (e) STED with corrected collar, (f) GB-STED. The insets in (a)-(f): Measured beam profile of depletion beam and cross section along y-axis at each depth. (g) Cross section of selected bead in (a)-(c), FWHM: 190 nm (STED without corrected

collar), 117 nm (STED with corrected collar), 116 nm (GB-STED); (h) cross section of selected bead in (d)-(f), FWHM: 218 nm (STED without corrected collar), 152 nm (STED with corrected collar), 121 nm (GB-STED). Scale bar: 500 nm. Dwell time: 0.4ms.

## 3. Preparation and characterization of phantom sample

Scattering and absorption are the key factors hindering high quality imaging in biological deep tissue. To validate the proposed GB-STED method, we performed imaging on phantom of gray matter in brain tissue. To prepare the phantom, we added the 40 nm-diameter beads in agarose with homogenized milk as scatter and India ink as absorber based on solid agarose. [7, 8] Figure S6 (a) shows the absorption coefficient of our agarose phantom with 2.5% homogenized milk and $2 \times 10^{-5}$ India ink at wavelength from 350 nm to 850 nm. At 671nm, the absorption coefficient is $\mu_a = 0.424 cm^{-1}$. Figure S6 (b) shows the setup to measure reduced scattering coefficient. A 671 nm laser diode was incident on the surface of agarose phantom at an angle of 45 degree and a CCD camera recorded the image of illuminated spot and scattered spot by two lens (lens1: F1=100mm, lens2: F2=125mm), shown in Figure S6 (c). The reduced scattered coefficient can be calculated as following equation: $\mu_s = \frac{\sin \alpha_i}{nD_{eff}} - 0.35 u_a$, [9] where $\alpha_i = 45°$ is induced angle, n=1.33 is refractive index of agarose phantom and $D_{eff} = D \times \frac{F1}{F2} = 0.512 mm$ . The calculated reduced scattered coefficient is $10.236 cm^{-1}$. The scattering and absorption properties are in agreement with those of gray matter in previous report. [10]

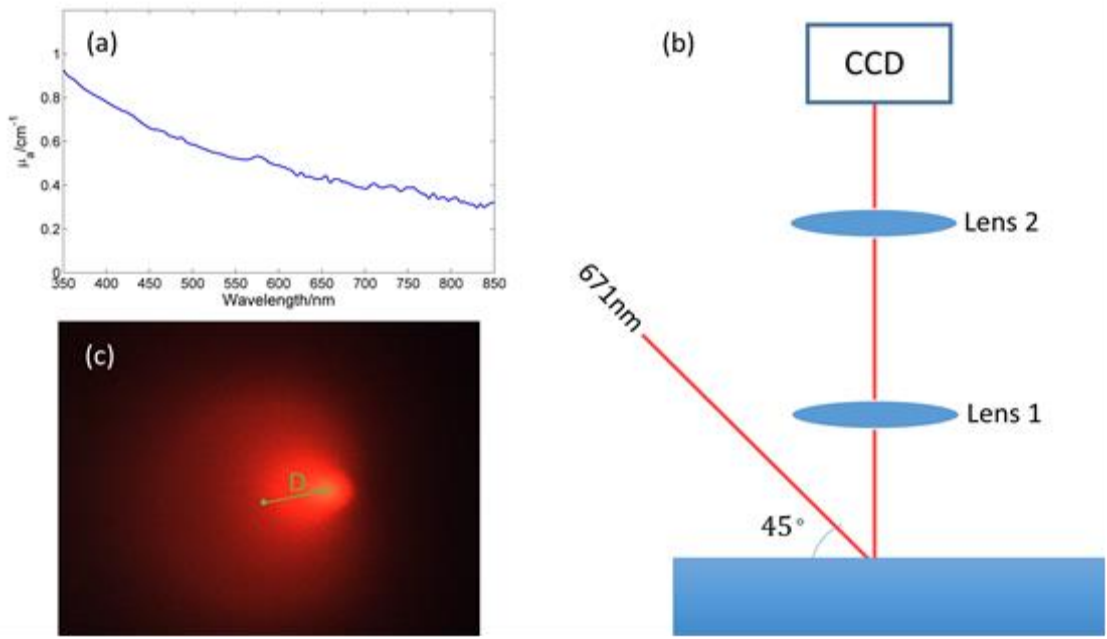

**Figure S6.** (a) Measured absorption coefficient of agarose phantom with 2.5% homogenized milk and $2 \times 10^{-5}$ India ink at wavelength from 350 nm to 850 nm. (b) The setup to measure reduced scattering coefficient, lens1: F1=100mm, lens2: F2=125mm. (c) The CCD image of illumination and scattering spots. The distance between the center of illumination spot and scattered spot is measured at $D = 0.64\text{mm}$.